\documentclass[12pt,preprint]{aastex}  

\received{March 27, 2006}
\slugcomment{To appear in Astrophys. J.}
\shorttitle{Heliosheath Interstellar Dust}
\shortauthors{Frisch et al.}

\newcommand{\commentpcf}[1]{}

\def\chipp{$\mu^\mathrm{''}$} 
\def\mgr{$m_\mathrm{gr}$} 
\def\Vgr{$V_\mathrm{gr}$} 
\def\Tgas{$T_\mathrm{gas}$} 
\def\lgyro{$l_\mathrm{gyro}$} 
\def\muG{$\mu $G}
\def\Upot{$U_\mathrm{ES}$}

\def\qtom{$Q/M$}
\def\glong{$\ell$}
\def\glat{$b$}

\def\Vlsr{V$_\mathrm{LSR}$}
\def\Javg{J$_\mathrm{avg}$}

\def\HH{H$_2$}

\def\HI{H$^{\rm o}$}

\def\HII{H$^{\rm +}$}
\def\HeI{He$^{\rm o}$}

\def\nHI{$n$(H$^{\rm o}$)}
\def\nHeI{$n$(He$^{\rm o}$)}

\def\nel{$n$(e)}
\def\Go{$G_\mathrm{o}$}
\def\Bis{$\stackrel{\rightarrow}{B_\mathrm{IS}}$}
\def\P{$P$}
\def\Pmax{$P_\mathrm{max}$}

\def\Bhat{$\hat{k_\mathrm{B}}$}

\def\absBis{$|{\stackrel{\rightarrow}{B_\mathrm{IS}}}|$}
\def\Bis{$\stackrel{\rightarrow}{B_\mathrm{IS}}$}
\def\Bhat{$\hat{k_\mathrm{B}}$}
\def\absBis{$|{\stackrel{\rightarrow}{B_\mathrm{IS}}}|$}

\def\Cpb{\hbox{$C_{\rm P \beta}$}}
\def\gl{\hbox{$l$}}
\def\gb{\hbox{$b$}}
\def\el{\hbox{$\lambda$}}
\def\eb{\hbox{$\beta$}}

\def\lambdazero{\hbox{$\lambda_\mathrm{0}$}}

\def\kms{\hbox{km s$^{-1}$}}
\def\deeg{\hbox{$^{\rm o}$}}
\def\NH{$N$(H)}

\def\HeI{He$^{\rm o}$}
\def\Halpha{H$\alpha$}
\def\cmtwo{cm$^{-2}$}
\def\thetac{$\theta_{\rm C}$}
\def\cc{cm$^{-3}$}

\def\Rgd{R$_\mathrm{g/d}$}

\begin{document}                                                                
\title{Interstellar Dust at the Magnetic Wall of the Heliosphere.  II.  }
\author{Priscilla C. Frisch}
\affil{Department of Astronomy and Astrophysics, University of
Chicago, Chicago, IL 60637.} \email{frisch@oddjob.uchicago.edu}

\begin{abstract}
\today

Two sets of data show that small interstellar grains captured in
interstellar magnetic fields, \Bis, draped over the heliosphere appear
to polarize the light of nearby stars.  The polarizing grains couple
to \Bis, while larger grains couple to the cloud velocity.  The
maximum polarization direction, \Pmax, is offset in ecliptic longitude by $ \Delta \lambda
\sim +35$\deeg\ from the upwind direction, and the polarization peak
is enhanced for stars near the ecliptic plane.  A band of weak
polarization stretches through the southern ecliptic hemisphere
between the \Pmax\ region and the downwind direction.  The \Pmax\
direction indicates that \Bis\ at the heliosphere is directed towards
\glong$\sim$105\deeg, forming an angle of $\sim$75\deeg\ with the
inflowing gas.  Grain alignment appears stable as the grains approach
the heliosphere.  The disruption of grain alignment in the tenuous
local interstellar material (ISM) by stochastic collisions is $\sim$600 
times slower than in the denser 
clouds for which grain alignment theory has been developed, however 
grain alignment mechanisms are still fast, and are boosted by compression 
of \Bis\ at the heliosphere.  The polarization vanishes where the outer
heliosheath magnetic fields become tangled or reconnect.  A similar
offset seen in energetic neutral atoms may form inside of the
polarization region, where the plasma is pinched by opposing magnetic
fields.  The composition of dust in the interstellar cloud surrounding
the solar system appears to be similar to olivines, and significant
quantities of carbonaceous grains can be ruled out.

\end{abstract}

\section{Introduction} \label{sec:intro}

The symmetry of the interaction between the heliosphere and
interstellar material is determined by three fundamental phenomena:
the solar wind (which loosely traces ecliptic coordinates and is
variable), the inflow direction of interstellar gas, and the
interstellar magnetic field.  This note is focused on the polarization
of interstellar dust grains at the heliosphere, which appear to mark
the third, and most poorly known, of these three parameters.

The heliosphere is the ``bubble'' occupied by the solar wind
\citep[e.g.][]{Zank:1999}.  The heliosphere nose location is provided
by the upwind direction of the flow of neutral interstellar gas
through the solar system.  Ulysses \HeI\ data provide the best value
for this velocity, which in the solar rest frame corresponds to --26.3
\kms\ from the upwind direction \el=254.7$\pm$0.5\deeg,
\eb=5.2$\pm$0.2\deeg\ \citep[ecliptic coordinates; in galactic
coordinates the upwind direction is \glong=3.3\deeg,
\glat=15.9\deeg,][]{Witte:2004,Moebiusetal:2004}.  Interstellar dust
grains have been observed flowing through the solar system by Ulysses
(at all ecliptic latitudes), Galileo, and Cassini
\citep[][F99]{Gruen22:1993,Baguhletal:1996,Gruenetal:2005,Frischetal:1999}.
The inflowing larger grains are at the gas velocity, and have an
upwind direction of \el=259\deeg, \eb=8\deeg\ (\gl=8.3\deeg,
\gb=12.9\deeg).  The Ulysses and Galileo data were acquired in the
1990's, during a time when the solar magnetic polarity `defocused'
smaller charged grains \citep[radius $a \lesssim0.2$
\micron,][]{Landgraf:2000}.  The termination shock (TS) of the solar
wind, where the solar wind becomes subsonic, was recently crossed by
Voyager 1 at $\sim$94 AU
\citep[e.g.,][]{Stoneetal:2005,Burlagaetal:2005,Deckeretal:2005}.
Beyond that $\sim$50 AU, the heliopause separates solar wind and
interstellar plasmas; outside of the heliopause, interstellar field
lines drapes over the heliosphere, following compression in the outer
heliosheath regions by factors of 2--5
\citep{Linde:1998,Pogorelovetal:2004}.

The primary data on the interstellar magnetic field near the Sun,
\Bis, have been observations of weakly polarized, \P$\sim$0.02\%,
light from stars within $\sim$30 pc and in the interval \glong $\sim
330^\circ \pm 60^\circ $, \glat$\sim 0 ^\circ \pm 50^\circ$
\citep[][T82, also see Frisch 1990]{Tinbergen:1982}.
\nocite{Frisch:1990} The polarization is produced by magnetically
aligned small charged interstellar dust grains, producing a
birefringent medium where the opacity parallel to \Bis\ is lower than
the opacity perpendicular to \Bis, resulting in a polarization maximum
when the sightline is perpendicular to the magnetic field direction,
\Bhat\ (\S \ref{sec:alignment}).  The location of the polarization
patch coincides with most of the mass of nearby interstellar material
\citep[ISM,][]{FrischYork:1983}, and with the upwind direction of the
bulk flow of ISM past the solar location \citep{FGW:2002}.

I recently discovered that this weak polarization coincides with the
heliosphere nose region, and suggested an origin associated with
magnetically aligned small grains trapped in the interstellar field
lines at the magnetic wall of the heliosphere \citep[][Paper
I]{Frisch:2005L}.  Paper I showed that the region of maximum
polarization is offset by $ \delta \lambda \sim +35\deeg$ along the
ecliptic plane from the large-grain inflow direction.
\emph{Evidently, at the heliosphere small grains with large
charge-to-mass ratios, \qtom, trace the interstellar magnetic field
direction, \Bis, while large grains with low \qtom\ trace the
interstellar cloud velocity.}

The present note provides support for this hypothesis that the
polarization originates with interstellar dust at the heliosphere,
using additional polarization data (\S \ref{sec:data}), evidence for
olivine grains with stable alignment as the grains approach the
heliosphere (\S \S \ref{sec:licdust}, \ref{sec:disrupt}), and a
discussion of the still-uncertain grain alignment mechanisms in the
context of the heliosphere interaction with the ISM (\S
\ref{sec:alignment}).

Charged ISDGs spin rapidly and will always be aligned so that the
observed polarization, \P, of starlight is parallel to the magnetic
field direction, \Bhat, regardless of the alignment mechanism
\citep[e.g.,][\S \ref{sec:alignment}]{Lazarian:2003}.  Evidently
aligned grains do not behave like dumb compass needles, but rather
trace the coupling between the grain angular momentum and \Bis.  Thus,
the T82 data indicate that \Bhat\ is approximately parallel to the
galactic plane and oriented towards \glong$\sim$90\deeg\ (see Fig. 
6 of T82), or as found here \glong$\sim$105\deeg.  A similar 
orientation for \Bhat\ is indicated by
the 2.6 kHz Langmuir emission events observed by Voyager
\citep[][KG03]{KurthGurnett:2003}.  Triangulation by Voyager 1 and 2
show that the dozen emission events detected in the 1990's arise in the
outer heliosphere at $\sim$100 AU from the Sun, and that these dozen
emission events are approximately aligned with the galactic plane
\citep[see, e.g.,][for a discussion of formation
mechanisms]{Cairns:2004}.

\section{Interstellar Gas, Magnetic Field, and Dust at Solar Location }\label{sec:lic}

\subsection{Gas}\label{sec:licgas}

The tenuous nature of the interstellar cloud surrounding the Sun
reduces the collisional disruption of grain alignment (\S
\ref{sec:disrupt}).  The density and ionization of the ISM at the
solar location are found from radiative transfer (RT) models that are
constrained by observations of ISM both inside and outside of the
heliosphere
\citep[][SF02,FS03,FS05,SF06]{SlavinFrisch:2002,FrischSlavin:2003,FrischSlavin:2005,SlavinFrisch:2006,SlavinFrisch:2006b}.
\footnote{The SF02 radiative transfer models are based on Version
90.05 of the CLOUDY code.}  While a range of tested equilibrium models
reproduce the general properties of tenuous clouds, the models which
provide the best fits to the local interstellar cloud (LIC) data, have
\nHI$\sim$0.2 \cc, \nel$\sim$0.1 \cc, \nHeI$\sim$0.015 \cc, and
fractional ionizations $\chi$(H)$\sim$0.29,
$\chi$(He)$\sim$0.47.\footnote{The fractional ionization of H is given
by $\chi$(H)=\HI/(\HI + \HII), etc.}  However somewhat lower
ionizations, \nel$\sim$0.07 \cc, can not yet be ruled out conclusively
(FS05).  The Ulysses \HeI\ data give a LIC temperature at the
heliosphere of 6300$\pm$340 K \citep{Witte:2004}.  If observed from a
distance, the warm partially ionized local ISM within $\sim$35 pc,
\NH$\lesssim 10^{19} $ \cmtwo, may appear as either warm ionized or
warm neutral material (WIM, WNM).

Interstellar dust grain abundances in generic warm ISM are relatively
invariant the ionization level of the gas, and approximately
25\%--30\% of the infrared emission from dust within $\sim$150 pc is
formed in warm \HII\ gas.  This is shown by observations of WNM and
WIM in a high-latitude region with sparse ISM, where comparisons are
made between 100---1000 \micron\ DIRBE and FIRAS infrared emission,
\HI\ 21-cm emission (Leiden data), and \HII\ \Halpha\ emission
\citep[WHAM data,][]{LagacheHaffneretal:2000}.  Column densities for
LIC-like clouds, \NH$ < 10^{18} $ \cmtwo, are not usually resolved in
\HI\ 21-cm or \Halpha\ measurements.

The LIC velocity, \Vlsr$\sim$15--20 \kms\ in the Local Standard of
Rest \citep[LSR,][]{FGW:2002}, is a factor in dust composition, since
high cloud velocities are associated with enhanced gas-phase
abundances of refractory elements and grain destruction in
interstellar shocks \citep{RoutlySpitzer:1952,Jones:1996}.  For
comparison, over 25\% of the \HI\ mass detected in the Arecibo
Millenium survey \citep{HTI} is at \Vlsr $>$10 \kms.  Local
interstellar gas, therefore, appears to be typical dust-containing
tenuous intermediate velocity partially ionized ISM.

\subsection{Magnetic Field}\label{sec:magnetic}

\citet{Linde:1998} has modeled the distance of the termination shock
(TS) in the direction of the Voyager 1 motion, and found that
\absBis$\sim$1.5 \muG\ yields a TS distance of 94 AU, which agrees
with the recently measured Voyager 1 value (\S \ref{sec:intro}).  I
argued that \Bis\ at the Sun should be similar to the uniform
component of \Bis\ inferred from pulsar data, \absBis$\sim$ 1.6 \muG,
which dominates low density interarm regions such as surrounding the
Sun \citep{Frisch:1990}.  However stronger fields are indicated ,
\absBis $\sim$2.6 \muG, if equipartition between thermal and magnetic
pressure applies.  The value \absBis=1.5 \muG\ will be used in the
following discussions.

\subsection{Dust Composition}\label{sec:licdust}

The interaction between ISDGs and the heliosphere depends on the dust
charge, mass, and composition.  In this section the results of
radiative transfer models of the local ISM are compared with a
reference abundance for the ISM, here assume to be solar abundances,
to determine the dust composition.  The following section gives the
gas-to-dust mass ratio calculated from the same assumptions.
Unfortunately, both solar abundances and the ISM composition,
generally assumed to be the summed abundances of the gas and dust, are
highly uncertain.  For instance, estimates of the solar ratio for O/H
vary by $\sim$30\%
\citep{GrevesseSauval:1998,Holweger:2001,Lodders:2003,Asplundetal:2005}.
Also, gas and grains may decouple in transient violent interstellar
phenomena \citep[e.g.][]{Slavinetal:2004}.

The predicted LIC gas-phase abundances of C, N, O, Mg, Al, Si, and Fe
are listed in Table \ref{tab:dust} for the best-fitting RT models 2
and 8 (SF02).  Comparisons between the gaseous Fe, Mg, Si, and O
abundances and solar abundances then yield underabundances of Fe, Mg,
Si, and O in the gas within $\sim$1 pc of the Sun towards $\epsilon$
CMa.  For this discussion, the abundances presented in
\citet[][L03]{Lodders:2003} are utilized.  The short length of the
local ISM towards $\epsilon$ CMa ($\sim$1 pc) contains two velocity
components separated by $\sim 8$ \kms\
\citep{GryJenkins:2001},\footnote{Radiative transfer models of the
LIC, alone, are now under development (SF06b).} but is characteristic
of tenuous ISM.

The atoms ``missing'' from the gas, and therefore inferred to be in
the dust, show approximately equal amounts of Fe, Mg, and Si, or
Fe:Mg:Si$\sim$1:1:1 (column 4 of Table \ref{tab:dust}).  The best LIC
dust constraints are the Mg:Si:Fe ratios, because the RT models
correct for \HII.  These ratios Fe:Mg:Si$\sim$1:1:1 suggest an
olivine-type mineral, possibly amorphous olivine similar to $\sim$85\%
of the dust mass towards SgrA$^*$ (see below).

Similar results are reached if the \citet{Asplundetal:2005} solar
abundances are used, except that there is a surplus of C in the gas
phase (see SF06), the O:Si ratios are reduced in the dust, and the
gas-to-dust mass ratio \Rgd\ becomes excessively large (\S
\ref{sec:rgd}).

An improved understanding of the reference abundance for local ISM
will permit stricter constraints on the LIC dust based on the ratio
O:Si.  The inferred O abundances from the models 2 and 8 indicate
somewhat higher O abundances in the dust than required by olivines,
since O:Mg, O:Si, and O:Fe are $\sim$5--6, rather than the value
$\sim$4 expected for olivines.  Evidently for olivine grains and L03
abundances, where O/H=575 PPM, between one and two O atoms are
``missing'' from both the dust and gas.  A value O/H=520 would be
required to eliminate these missing O atoms.

The LIC grains are not carbonaceous.  The RT models 2 and 8 give C
abundances that are approximately solar, suggesting that graphite is
absent from LIC dust, perhaps because it did not survive the LIC
acceleration mechanism \citep[Table \ref{tab:dust}, also
see][]{SlavinFrisch:2006}.  The absence of local graphite does not
affect the polarization discussion, since graphite does not contribute
to polarization in the global ISM \citep{Whittet:2004}.

Silicates are a widespread constituent of interstellar dust.  A
silicate mixture consisting of 84.9\% amorphous olivine (MgFeSiO$_4$)
and 15\% amorphous pyroxene (MgFeSi$_2$O$_6$), by mass, fits
observations of a 9.7 \micron\ infrared feature in Sgr A$^*$
\citep[][]{Kemperetal:2004}.  The amorphous olivine grains forming
85\% of the grain mass towards SgrA* must be robust, or they would not
dominate, which lends credibility to the possibility that local ISDGs
are amorphous olivines that have survived the process that accelerated
the moderate velocity ISM surrounding the Sun.  Silicates are also
found in evolved stars such as Mira variables
\citep[e.g.][]{Dorschner:1995,ChiarTielens:2006}.  Additional evidence
of widespread silicates are provided by models of the grain
populations, in terms of size and composition, required to fit
extinction measurements, diffuse infrared emission, and abundance
constraints for the diffuse ISM \citep[][ZDA]{ZubkoDwekArendt:2004}.
The ZDA COMP-GR-FG model, consisting of silicates (MgFeSiO$_4$),
graphite, and a small amount PAHs, provides a good fit to the
extinction and IR emission data if ISM abundances are comparable to F
and G star abundances.  The silicates in the ZDA COMP-GR-FG model have
$a < 0.25$ \micron, with most of the grain mass contained in grains
with $a \sim 0.06 - 0.25$ \micron, however a significant fraction of
silicate grains with $a \lesssim 0.01$ \micron\ are present.  The
small amount of PAHs required by the COMP-GR-FG are allowed by the LIC
C data, depending on the assumed solar C abundance.

\subsection{Gas-to-Dust Mass Ratio, \Rgd\ }\label{sec:rgd}

The above arguments used to derive grain composition assume that dust
and gas have been fully mixed over the cloud lifetime.  A check on
this assumption is the gas-to-dust mass ratio, \Rgd, in local ISM,
which can be examined in several ways.  (1) The first method compares
spacecraft $in~situ$ measurements of the ISDG mass with the gas mass
determined from RT models.  Updating the F99 discussion with more
recent spacecraft results \citep{Landgrafetal:2000,Altobellietal:2004}
yields \Rgd$<$130, where the upper limit holds because small charged
ISDGs are excluded from the solar system.  (2) \Rgd\ can also be
determined from comparisons between the RT models and solar
abundances.  For the L03 solar abundances, model 8 yields \Rgd=186.
If instead the ISM reference O abundance is O/H=520 (see above), model
8 gives \Rgd=210.  For either case the $in~situ$ measurements and
model 8 predictions disagree.  For comparison, the Asplund et
al. abundances yield \Rgd=330, due to the lower solar abundances for
Mg, Si, and Fe of $\sim$20\%, compared to L03.  The difference between
(1) and (2) estimates of \Rgd\ may indicate a grain population is
present that has not been coupled to the gas over the cloud lifetime,
such as expected if the LIC has been shocked to high LSR velocities
\citep[][F99]{Frisch:1981}.  Evidently the tiny grains are
disproportionally destroyed by sputtering so that \Rgd\ derived from
$in~situ$ data may not be significantly overestimated.

A third basis for understanding \Rgd\ in the LIC uses the silicate
component of the ZDA COMP-GR-FG model as a model for LIC dust.  The
silicate component in the COMP-GR-FG model alone yields \Rgd=251,
while the silicate and PAHs together give \Rgd=233.  The other
constituents in the COMP-GR-FG model, graphite, water ice, and
organics, are not required by ``missing'' atoms of RT models 2 and
8. If allowance is made for differences between the F and G star
abundances of ZDA and the L03 abundances assumed here, and the F and G
star abundances are compared to model 8, then the \Rgd=229 for model 8
is similar to the COMP-GR-FG predictions for the silicate and PAHs
combined.

The above arguments indicate that the local ISM gas phase abundances
given by RT models 2 and 8 of SF02 are consistent with the ZDA
COMP-GR-FG silicate grains, providing that graphite is absent from the
LIC.  The silicate grains in the ZDA COMP-GR-FG model have $a <0.4$
\micron, compared to $in~situ$ LIC data showing grains up to $a
\gtrsim$1.1 \micron\ (F99).  Disproportionate percentages of large
grains are predicted by the \citet{Slavinetal:2004} grain destruction
models for regions inside of supernova remnants, where the large
grains survived multiple shocks associated with such environments.
This scenario agrees with models of the LIC as a fragment of a
superbubble shell, forming $\sim$4 million years ago out of debris
left over from a prior generation of star formation in the
Scorpius-Centaurus Association
\citep[][]{Frisch:1981,Frisch:1995,Frisch:1996}.\footnote{Please note that \citet{Breitschwerdtetal:2000}
were incorrect when they stated ``...an alternative
scenario by Frisch (1995, 1996), ... has pictured the Local
Bubble to be an appendix of the Loop I superbubble which expanded
into a low density interarm region, an idea that had been proposed
earlier by \citet{Bochkarev:1987}.  ''.  I proposed this 
idea in 1981, and Bochkarev referenced my earlier paper.}

\section{Characteristics of the Polarization }\label{sec:data}

The analysis of Paper I has been repeated with the inclusion of
polarization data from \citet{Piirola:1977}, as well as the
\citep{Tinbergen:1982} data.  (Paper I was based only on the Tinbergen
data.)  These combined data sets yielded a sample of 202 nearby stars;
86\% of the stars are within 40 pc.  Note that these distances are
based on Hipparcos data, which may differ from distances in the
original pre-Hipparcos publications.  The two data sets have 45 stars
in common.  Both data sets were acquired during the solar minimum
conditions of 1973--1974,\footnote{The Piirola data were acquired in
1974 \citep{Piirola:1977}, while the Tinbergen data were acquired
during 1973 (J. Tinbergen, private communication).}  when the solar
magnetic polarity was north pole positive \citep[i.e. A$>$0, field
lines emerging at the north pole,][]{Frischetal:2005}.  The typical
1$\sigma$ measurement uncertainties for these data are $\sim 6-17
\times 10^{-5}$ degree of polarization (Channels I and II), but only
values with uncertainties $\lesssim 7 \times 10^{-5}$ are used here.
The linear Stokes parameters U and Q provided in the original papers
give \P\ and the polarization position angle, \thetac=0.5 arctan(U/Q),
which is the angle of the observed electric vector as measured
positive in a counter-clockwise direction from the north celestial
pole.  The position angle, \thetac, indicates a value in the celestial
coordinate system.

\begin{figure}[t]
\plotone{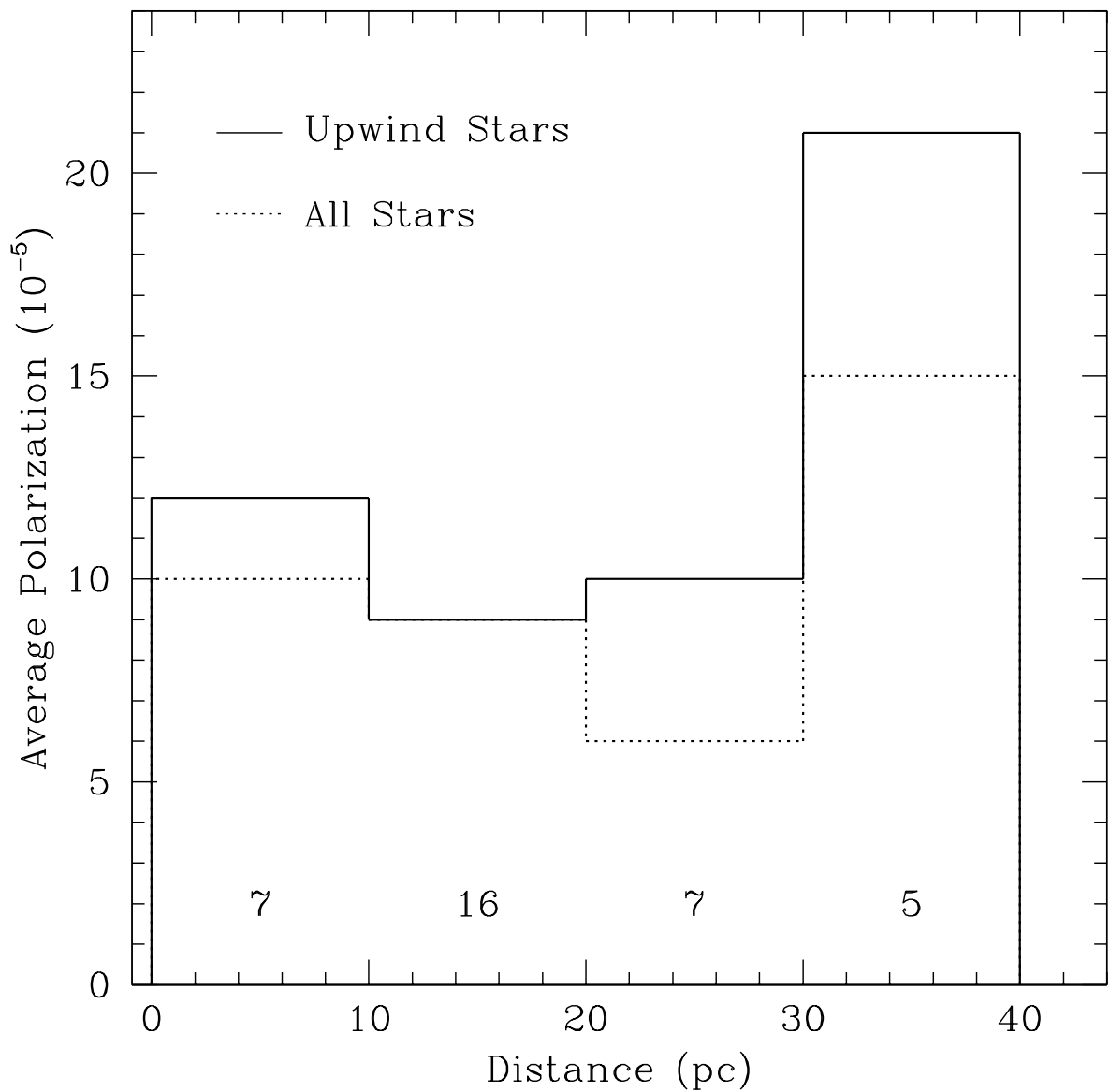}
\caption{Histogram of the average polarization for stars
binned into distance intervals of 10 pc.
The polarization units are $10^{-5}$ degree of polarization, and 
1$\sigma$ uncertainties are $\sim  7 \times 10^{-5}$.
The solid line includes only stars towards the heliosphere nose
($330^\circ < \ell < 40^\circ$ and $|b| < 50^\circ$), with the
numbers of stars in each bin shown at the bottom.
The dotted line includes all stars within 40 pc.  
\label{fig:f1} }
\end{figure}

The primary results of Paper I are shown in Fig. \ref{fig:f1}, but now
including both \citet{Tinbergen:1982} and \citet{Piirola:1977} data.
The three panels show the ecliptic longitude, \el, dependence of
polarization \P\ (top), position angle \thetac\ (middle, \thetac\ in
celestial coordinates), and the \P-\eb\ correlation coefficient (\Cpb,
bottom, \eb\ is ecliptic latitude).  The filled dots represent values
averaged over stars within $\pm$50\deeg\ of the ecliptic plane ($ |
\beta | < 50^\circ $), and within $\pm$20\deeg\ of the central
ecliptic longitude, \lambdazero.  An artificial smoothing of the
points is introduced by the 3\deeg\ stepping interval of \lambdazero\
along the ecliptic plane, but this smoothing is tolerated because of
the relatively small sample size.  Stars within 20\deeg\ of the
ecliptic plane ($| \beta | < 20^\circ$) were analyzed separately and
are plotted as dashed lines in the top two panels.  \Cpb\ represents
the covariance between polarization \P\ and ecliptic latitude $\eb$
(eqn. 1 of Paper I).  Fig. \ref{fig:f1} also shows the direction of
the large inflowing ISDGs observed by Ulysses and Galileo (\S
\ref{sec:data}) The polarization maximum seen near
\lambdazero$\sim$140\deeg, with $\sim$1.8 $\sigma$, is dominated by
the three stars HD 98230, HD 78154 and HD 95128 at high ecliptic
latitude.  The conclusions of Paper I are reinforced by this extended
data set.

\subsection{Distance Dependence of Polarization} \label{sec:distanceP}

The mean polarization is independent of star distance for stars within
$\sim$30 pc, but increases by $\sim 1 \sigma$ beyond 30 pc
(Fig. \ref{fig:f2}).  In contrast, the polarizations of the total
sample of $\sim$200 stars show no enhancement in the nearest 10 pc,
and barely increase beyond 30 pc.  There does not appear to be any
significant change with distance in \thetac\ for the stars
contributing to the \Pmax region, however the statistics are poor
since these are $\sim 2-3 \sigma$ effects.

The nearest star showing polarization is 36 OphAB (HD 155886) at a
distance of 5 pc, and located at \glong=358.3\deeg, \glat=+6.9\deeg\
(\el=260.0, \eb=--3.5\deeg).  This star is close to the ecliptic
plane, 10\deeg\ from the \HeI\ upwind direction, and with \P$= 1.75
\pm 0.7 \times 10^{-4}$ and \thetac=--15\deeg.  Given the uncertainties 
and projection effects, this is consistent with $<$\thetac$> \sim -23^\circ$,
averaged over the stars in the \Pmax\ region (see the following section).
A 1$\sigma$ uncertainty in either Q or
U translates to an uncertainty in the position angle of
$\delta$\thetac=3\deeg--8\deeg\ for 36 OphAB.  The relatively constant
\P\ value for the nearest 30 pc, of which 36 OphAB is an example,
indicates the polarization is formed very close to the Sun.  

\subsection{Direction of Maximum Polarization} \label{sec:pmax}

The polarization reaches a maximum, \Pmax, at a position offset by
$\Delta \lambda \sim +35 ^\circ$ from the heliosphere nose given by
the large grains and \HeI\ (Fig. \ref{fig:f1}, top).  The polarization
maximum is strongest for stars close to the ecliptic plane, $| \beta|
< 20 ^\circ$ (dashed line).  Considering both \P\ and \thetac\ in
Fig. \ref{fig:f1} indicates that the maximum polarization occurs at
\el=294\deeg$\pm$7\deeg\ and $ | \beta | < 20^\circ $
(\glong$\sim$19.2\deeg, \glat$\sim$--21.2\deeg) with an uncertainty of
$\sim$20\deeg.  Fig. \ref{fig:f3} indicates the maximum is centered
around \el=294\deeg, \eb$\sim -5^\circ$, or \glong$\sim$14.6\deeg,
\glat$\sim$--23.8\deeg.  Presumably the polarization is no longer
observable when the angle between the sightline and \Bhat\ deviates
significantly from 90$^\circ$.  This condition is easy to estimate,
since the observed polarization is very weak and generally not much
more than $\sim$3.0 $\sigma$, while the minimum reliable polarization
measurement is $\sim$2.5 $\sigma$.  Thus polarization will only be
detectable in stars within $\sim 35^\circ$ of the perpendicular
direction if the field lines are straight.  The width of the band of
observed polarization in Fig. \ref{fig:f3} is $\sim$45 \deeg, which is
not inconsistent with this estimate considering the uncertainties.

\subsection{Polarization Position Angle} \label{sec:thetac}

The consistency of \thetac\ in the \Pmax\ region contributes to the
reality of this weak polarization signal.  The polarization position
angle \thetac\ has a consistent value of \thetac$\sim$--23 \deeg\ in
the region of maximum polarization (Fig. \ref{fig:f1}, middle).  Note
that 36 OphAB is $\sim$35\deeg\ from the \Pmax\ direction, but retains
a consistent polarization angle.  If the \Pmax\ region corresponds to
a sightline perpendicular to \Bis, as expected, then
\Pmax$\sim 20 \times 10^{-5}$ projects to \Pmax$\sim 17 \times 10^{-5}$
 for the 36
OphAB direction, versus the observed value \thetac$\sim$--15\deeg,
which is within the uncertainties.  Therefore, the position angles are
consistent with \Pmax\ corresponding to a sightline perpendicular to
\Bis.  For the \Pmax\ region centered at
\el$\sim$294\deeg, \eb=--5\deeg, and noting that the plane of
polarization is approximately parallel to the galactic plane (T82),
then \Bis\ within 5 pc of the Sun is directed towards
\glong$\sim$105\deeg, \glat$\sim$0\deeg, with an uncertainty of $\sim
\pm ~ 10$\deeg.

\subsection{\P--\eb\ Anticorrelation and Magnetic Wall }

In the \Pmax\ region, \el=280\deeg--330\deeg, polarization strength
anticorrelates with ecliptic latitude (Fig. \ref{fig:f1}, bottom).
This conclusion is unchanged from Paper I, and is based on the full
\eb=$\pm$50\deeg\ data set.  The anticorrelation between \P\ and \eb\
for \eb$<$0\deeg\ is the primary evidence that the polarization is
created by aligned grains in the magnetic wall of the heliosphere.
The magnetic wall is predicted to be at negative ecliptic latitudes
during the 1970's for an interstellar field directed towards
$\sim$90\deeg, when the solar wind and interstellar fields were
parallel in the southern ecliptic hemisphere.  The \Bis\ field
polarity is obtained from rotation measure data
\citep[e.g.][]{RandKulkarni:1989}.  Polarizations of stars within
$\sim$400 pc and pulsar rotation measure data yield similar
orientations for the global nearby \Bis, $\sim$83\deeg\ and
$\sim$71\deeg\ respectively; the rotation measure data also give the
field polarity, and global \Bis\ near the Sun is directed towards
\glong$\sim$71\deeg\
\citep{HeilesMF:1996,RandKulkarni:1989,IndraniDeshpande:1999}.  A
similar polarity is assumed for the LIC \Bis, yielding parallel solar
wind and interstellar magnetic fields in the southern ecliptic
hemisphere, during the solar minima of the 1970's and 1990's, when the
polarization and Ulysses data were acquired.  For \Bhat\ directed
towards \glong$\sim$105\deeg, during the 1970's the solar wind and
interstellar field lines were antiparallel over the northern ecliptic
hemisphere, so that reconnection would tangle the field lines and
rapidly disrupt grain alignment.

The polarization data are plotted in ecliptic coordinates in Fig.
\ref{fig:f3}, along with the positions of the galactic
plane, 2.6 kHz emission events (KG03, \S \ref{sec:intro}), the upwind
direction of the large inflowing dust grains (F99), and the upwind
directions for \HI\ and \HeI\ \citep[W04,][]{Lallementetal:2005}.
Polarizations with \P$> 2 \sigma$ are plotted as filled symbols.  The
stars with the strongest polarization form a wide band ($>$45\deeg)
extending from the upwind to downwind direction.
One side of the band extends through
the points (\el, \eb)$\sim$(270\deeg, --5\deeg), (360\deeg,
--15\deeg), to (85\deeg,--40\deeg), and the second side dips to
lower ecliptic latitudes and through (\el,
\eb)$\sim$(280\deeg, --45\deeg), (360\deeg,--50\deeg), to (40,
--60\deeg).  This band of weak polarization in the southern ecliptic
hemisphere is offset towards the direction of solar rotation, when
compared to a meridian passing through the upwind direction and south
ecliptic pole.

\section{Grain Alignment in Local Cloud }\label{sec:alignment}

Most observed starlight polarization is formed in denser ISM, however
the LIC grain composition appears similar to the olivines in the
global ISM (\S \ref{sec:licdust}) so that existing alignment theory
can be applied.  Magnetic alignment theories derive from the original
discussion of \citet[][DG51]{DavisGreenstein:1951},\footnote{Davis is
the same Leverett Davis, Jr., who originated the idea of the
heliosphere by proposing that the momentum flux of solar corpuscular
radiation formed a cavity of radius $\sim$200 AU in the local galactic
magnetic field \citep[also see][]{Parker:2001}.} who proposed that
magnetic torques align the principal inertial axis of the grain with
the average angular momentum of asymmetric grains, \Javg, and also
align \Javg\ with the magnetic field, \Bis.  Grain-gas collisions spin
up grains so they acquire a random angular momentum with two parts,
\Javg\ which precesses about \Bis, and a stochastically varying
angular momentum component related to the nutation of the asymmetric
grain body inertial moment about \Javg.  Torques both dissipate the
component of \Javg\ perpendicular to \Bis, and dampen grain nutation
about \Javg.  Uncertainties in the timescales for the magnetic
alignment of grains relates primarily to these poorly understood, but
heavily researched, torques
\citep[e.g.,][]{DavisGreenstein:1951,Martin:1974,Purcell:1979,LeeDraine:1985,Mathis:1986,LeeDraine:1985,RobergeLazarian:1999,Lazarian:2000,Draine:2003,Lazarian:2003}.
If grain alignment in the global ISM is rapid and robust
\citep{Lazarian:2003}, it will be even more robust in the LIC.  I
assume that ISDGs arrive at the heliosphere prealigned, so the
outstanding question becomes whether these alignments can be
randomized on relatively short time scales.

\citet{Lazarian:2003} discusses polarization by very small grains, $a
<< 0.06$ \micron; such grains would be tightly coupled to \Bis\ with
gyroradii $<$10 AU (F99).  The LIC does not contain \HH, so alternate
processes would be required to generate any suprathermal rotation
required to stabilize grain alignment against collisional disruption
\citep{Purcell:1979,Draine:1996}.  However, the disruption of grain
alignment in the tenuous LIC occurs over timescales longer by factors
of $\sim$600 than in denser clouds (see below), so that suprathermal
rotation is less important.

Grains are assumed to be silicates with density 3 gr \cc\ and a nearly
spherical geometry. The LIC densities, temperature, and magnetic field
strength are \nHI$\sim$0.2 \cc, \nel$\sim$0.1 \cc, $n$=\nHI+\nel,
\Tgas$ \sim$6300 K, and \absBis$\sim$1.5 \muG\ (\S \S
\ref{sec:licgas},\ref{sec:magnetic}).  \citep[also
see][]{GoodmanWhittet:1995}.  \citet{Mathis:1986} argues that grains
with radii $a=0.06 - 0.10$ \micron\ cause the polarization, based on
the wavelength of maximum polarization, extinction and abundance
constraints.  The lower size limit originates with the requirement the
grains are large enough to contain a superparamagnetic (SPM) inclusion  
(imaginary magnetic susceptibility \chipp$>$0).  The SPM inclusion
acts to increase the strength of the torque dissipating the grain body
angular momentum, so that the average angular momentum vector \Javg\
and largest inertial axis co-align with each other and \Bis\
\citep{DavisGreenstein:1951,Purcell:1979,Mathis:1986}.  Grains in this 
size range are excluded from the heliosphere for an initial \Bis$>$1.5
\muG, and as field strengths rise as the grains approach the magnetic
wall of the heliosphere.

\subsection{Stability of Grain Alignment in Low Density ISM }\label{sec:disrupt}

There is a rich body of literature, stretching over 50 years, that associates
interstellar polarization with asymmetric ISM opacities caused
by magnetically aligned interstellar dust grains.
However an important point to realize is that the LIC is warmer by a
factor of $\sim$100, less dense by a factor of $\sim$65, and has
plasma densities a factor of $\sim$10 larger, than the typical clouds
that have been discussed in the polarization literature.  If the LIC
and dense cloud magnetic field strengths are comparable to the ordered
and random components of the global \Bis\ \citep[$\sim$1.6 \muG\ and
$\sim$5 \muG\ respectively,][]{RandKulkarni:1989}, then the magnetic
field strength in the LIC is also a factor of $\sim$4 weaker than the
field strengths assumed in most polarization theory studies.
Collisions between gas and grains provide the primary means of
disrupting grain alignment.  In the LIC, with density $n \sim 0.3$ \cc\
and temperature $T \sim 6300$ K, the rate of these collisions is down
by a factor of $\sim n / \sqrt{T} \sim 600 $ compared to denser
clouds.  \emph{Therefore, even though \Bis\ is weaker, it is
fundamentally easier to magnetically align ISDGs in the LIC than in
denser clouds.}  The following discussion assumes that grains are
magnetically aligned in the upwind gas, and arrive prealigned to the
heliosphere vicinity with perhaps some alignment enhancement as \Bis\
is compressed at the heliosphere.

Several arguments indicate that stochastic processes do not affect
grain alignment in the heliosphere-ISM interaction region.  Grain
alignment in the LIC will be disrupted at least as often as grains
accumulate their own mass in thermal collisions with the gas, which
occurs over timescales of of $\tau_\mathrm{mass} \sim10^6 a
\rho_\mathrm{gr} / n \sqrt{T_\mathrm{gas} }$ Myrs = 0.7--1.2 Myrs.
The rotational damping time is $\tau_\mathrm{d} \sim 0.6
\tau_\mathrm{mass}$ for symmetric grains \citep[in this case
cubical,][]{LazarianDraine:1997}.  These timescales are long compared
to the $\sim$30 years required for a dust grain to cross the bow shock
and approach the heliopause.

In contrast, torques that dampen grain body nutation about \Javg\
occur over short timescales.  Elastic collisions between thermal gas
particles and spherical grains with $a \sim 0.1$ \micron\ radius will
spin the grains about the grain-body principal axis at a mean angular
velocity of $\omega \sim 3 \times 10^{4}
\sqrt{T_\mathrm{gas}}/\sqrt{a^{5}_{\mu} \rho} \sim 1.3 \times 10^6$ s$^{-1}$.
This ``thermal'' rotation induces a magnetic suseptibility of
K($\omega$)=\chipp/$\omega \sim 10^{-13}$ s, which can be enhanced to
$\sim 2 \times 10^{-8}$ s if 1\% of the grain material is contained in
SPM clusters with $\sim $5000 Fe atoms \citep[][]{Draine:1996}, and
which shortens the magnetic damping time to $\sim$5 years
\citep{RobergeLazarian:1999}.  The magnetic moment induced by the
rotating grain charge is also enhanced by the Barnett effect, where
grain body angular momentum is dissipated by interactions with the
grain lattice, and which dampens grain nutation and increases magnetic
alignment efficiency.  \citep[][]{Lazarian:2003,RobergeLazarian:1999}.  
Therefore, apparently stochastic mechanisms do not disrupt the alignment 
of LIC ISDGs arriving at the heliosphere, and for the most optimal 
assumptions alignment may increase on time scales of $\sim$5 years when
interstellar field lines are compressed at the heliosphere.

\subsection{Grain Charge and Gyroradius }\label{sec:gyro}

The outer heliosheath regions filter out small charged grains.  As
ISDGS approach the heliosphere, grains with small charge-to-mass
ratios, $Q$/\mgr\ follow the neutral gas into the heliosphere.  Grains
with large $Q$/\mgr\ ratios, together with interstellar plasma
components, are trapped by \Bis\ as field lines drape over the
heliosphere \citep[F99,][]{KimuraMann:1998}.  During this interaction,
grain gyroradii, \lgyro, vary with \Bis\ and grain charging rates.
For grain velocity \Vgr, \lgyro=\mgr \Vgr/$e Z$\absBis, where $Z = 4
\pi \epsilon_o a$\Upot\ is a function of the electrostatic grain
potential \Upot, and $\epsilon_o = 8.9 \times 10^{-14}$ C V$^{-1}$
cm$^{-1}$ is the permittivity (F99).  \Upot\ is determined partially
by photoelectric emission rates and collisional grain charging.  In
the LIC, \Upot$\sim 0.7-1.4$ V for $a=0.06 - 0.1$ \micron\
\citep[e.g.][]{WeingartnerDraine:2001a}.
For \Vgr=26 \kms, $a \sim 0.1$ \micron, and \absBis=1.5 \muG, \lgyro$
\sim 130-175$ AU; for $ a \sim 0.06$ \micron, \lgyro$\sim 35-50$ AU.
The factor of $\sim$2 jump in \absBis\ at the bow shock, and again
near the heliopause, decreases \lgyro\ \citep{Linde:1998}.  The
self-consistent treatment in F99 of grain charging and \Bis\
compression at the heliosphere indicates that for \absBis=1.5 \muG\
and radiation field parameter \Go$\sim$0, grains with $ a
\lesssim$0.06 \micron\ are excluded at the bow shock, and grains with
$ a \lesssim$0.15 \micron\ are excluded at the heliopause.  Grains
with radii $a < 0.01$ \micron\ are excluded from the heliosphere under
nearly all conditions.  Analogous results were found by
\citet{KimuraMann:1998} for strong fields, \Bis$\sim$13 \muG, such
that trajectories of grains with $a < 2$ \micron\ were found to be
altered upwind of the heliopause.  Grains in the size range of the
Mathis polarizing grains, 0.06--0.1 \micron, are excluded from the
heliosphere.

\subsection{Slowly Changing \Bis}\label{sec:bvar}

In general, for magnetically aligned ISDGs the polarization maximum is
observed where the sightline is perpendicular to the field direction.
However, as the LIC interacts with the region beyond the heliopause,
either grain alignment does or does not follow the slowly changing
direction of \Bhat.  \citet{Lazarian:2003} argues that the fast Larmor
precession, $\tau_\mathrm{Lar}$, induced by the Barnett effect
proceeds at a period of $\tau_\mathrm{Lar} \sim 15 / B_\mathrm{IS}$ s
$\sim$0.3 years for \absBis=1.5 \muG.  If this short timescale truly
characterizes the alignment adjustment of grains in a slowly varying
\Bhat, then the maximum polarization direction will trace the
perpendicular of the local field direction, rather than the ``fossil''
perpendicular direction indicated by grains that do not realign
quickly enough beyond the heliopause.  The transfer properties of the
Stokes parameter Q over distances that are small compared to
variations in \Bhat\ indicates that weak polarization is weakly
enhanced, and this case should apply near the heliosheath
\citep{NeeJokipii:1979}.

\section{Discussion and Conclusions}  \label{sec:conclusions}

The composition of LIC dust can be found from radiative transfer
models, and the assumption that a solar abundance's worth of atoms are
either in the dust or the gas.  The resulting LIC dust composition
indicates olivines, that are perhaps similar to the amorphous olivines
that dominate the SgrA* sightline (\S \ref{sec:licdust}).  Significant
quantities of carbonaceous grains can be ruled out.  These grains give
a gas-to-dust mass ratio of \Rgd$\sim$200, which is greater than the
$in~situ$ value of \Rgd$<$130 (\S \ref{sec:rgd}).  This indicates that
the gas and dust may have decoupled over the LIC cloud lifetime, which
in turn negates the underlying assumption that the reference abundance
for an interstellar cloud is constant over small spatial scales.
This contradiction may arise from the poorly understood reference
abundance for the ISM.

The small interstellar grains causing the polarization, with high
\qtom, trace \Bis\ at the heliosphere, while large grains with low
\qtom\ couple to the LIC velocity.  The gyroradii of grains
interacting with the heliosphere in the presence of a weak \Bis\ has
been studied previously, and grains with radii 0.06--0.10 \micron\ are
excluded, \lgyro$<$50 AU, as the interstellar field compresses against
the heliosphere.

An offset of $ \Delta \lambda \sim +35$\deeg\ between the direction of maximum
polarization \Pmax\ and the upwind direction, found in Paper I, is
confirmed with the addition of the Piirola data (\S \ref{sec:data}).
Evidently this offset does not act like a dumb compass needle, since
grain alignment traces the coupling between the grain angular momentum
and \Bis.

A band of weak polarization, $\sim 2 \sigma - 3 \sigma$, stretchs through
the southern ecliptic hemisphere between the \Pmax\ region and the
downwind direction.  The width of the band in the upwind direction is
consistent with the detectability limits of the polarization
(\S \ref{sec:pmax}).  The band shows an offset towards the
direction of positive solar rotation when compared to a meridian
between the upwind direction and south ecliptic pole.

Consistent position angles for polarizations in the upwind direction,
and the relative constancy of polarization with star distance,
indicate that that polarization is formed within 5 pc of the Sun.
Polarization data are also consistent with \Pmax\ occurring in a
sightline perpendicular to the interstellar magnetic field direction,
\Bhat, because the polarization towards 36 OphAB is decreased by the
cosine of the angle between the star and \Pmax\ direction.  

The position angles of the polarization in the \Pmax\ region are
relatively parallel to the galactic plane (\S \ref{sec:thetac}).
Together with the \Pmax\ direction of \glong$\sim$15\deeg, it then
appears that \Bis\ is oriented towards \glong$\sim$105\deeg,
\glat$\sim$0\deeg\ with $\sim \pm 10^\circ$ uncertainties.  This
conclusion makes use of grain alignment theories that indicate the
plane of polarization is parallel to \Bis.  If the polarity of the LIC
\Bis\ is similar to the polarity of the global field found from
rotation measure data, then the field is directed towards
\glong$\sim$105\deeg.  In this case, the interstellar and solar wind
magnetic fields were parallel in the southern ecliptic hemisphere
during the solar minimum of the 1970's when the polarization data were
acquired, and again during the 1990's minimum when much of the Ulysses
dust data were obtained.  This \Bhat\ direction gives an angle between
the inflowing neutral gas and \Bis\ of $\sim$75\deeg.

In low density ISM such as the LIC, the disruption of grain alignment
is $\sim$600 times slower than in the denser clouds for which grain
alignment theory has been developed (\S \ref{sec:disrupt}).  It takes
on the order of $10^6$ years for grain alignment to disrupt by
sweeping up its own mass in collisions.  The ISDGs are assumed to
arrive prealigned at the heliosphere, and collisional disruption of
alignment in the LIC is too slow to be meaningful.

Despite the fact that grain alignment theories are still somewhat
uncertain, collisions between gas particles and grains occur rapidly
enough to allow magnetic torques to quickly dissipate the nutation of
the grain body moment of inertia about the average angular momentum,
\Javg, so that grain alignment is quick and robust even in low density
ISM.  Estimates place the alignment on the order of years because of
the Barnett effect (\S \ref{sec:alignment}), so the alignment of
grains near the heliosphere will respond rapidly to the slowly varying
direction of \Bis.  If \Bis\ becomes tangled or reconnects with the
solar wind magnetic field, as happened in much of the northern
ecliptic hemisphere during the 1970's solar minimum according to this
\Pmax\ direction, then the polarization should vanish.  The
compression of the interstellar magnetic field in the outer
heliosheath regions, by a factor of $\sim$5, boosts grain alignment.
During solar minimum conditions the solar magnetic field is
dominated by open field lines with a dipole component, in
contrast to solar maximum conditions when the quadrapole and hexapole
moments dominate, making the solar wind magnetic field
in the outer heliosphere more disorderly \citep{Bravoetal:1998}.

Polarization will vanish where the field direction changes rapidly, is
turbulent over smaller scales than the grain gyroradius, or where the
interstellar and solar wind magnetic field are anti-parallel.

\begin{figure}[h!]
\begin{center}
\includegraphics [height=6.2in,width=6.2in]{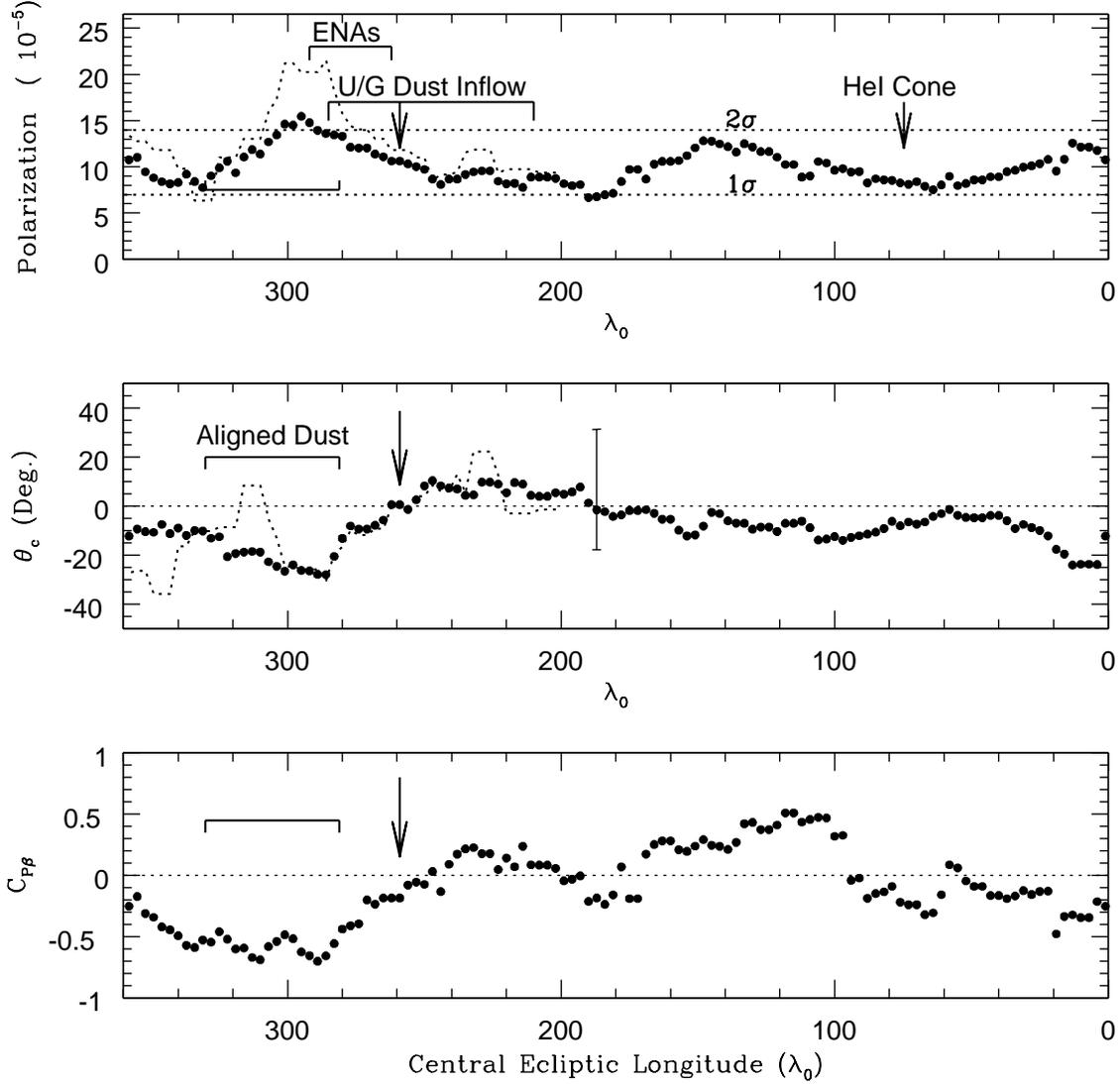}
\end{center}
\caption{ Polarization Properties as a Function of Ecliptic Latitude:
{Top:} The average polarization \P\ for stars with $| \beta|
<$50\deeg\ is plotted as dots, and for stars with $| \beta| <$20\deeg\
as a dashed line.  Data are averaged over $\pm 20$\deeg\ in ecliptic
longitude, $\lambda$.  The direction of maximum \P\ is shifted by
$\sim$20--30\deeg\ from the upwind direction of the large interstellar
dust grains detected by Ulysses/Galileo \citep{Frischetal:1999}, and
upwind gas and dust directions differ by $\sim$5\deeg.  {Middle:} The
averaged polarization position angle \emph{in celestial coordinates},
\thetac.  In the region of maximum polarization, $\lambda \sim
280^\circ \rightarrow 310^\circ$, the grains show consistent position
angles.  {Bottom:} The correlation coefficient between \P\ (top) and
$\beta$ is shown as a function of the ecliptic longitude.  The
strongest polarization is found at negative ecliptic latitudes.
\label{fig:f2} }
\end{figure}

\begin{figure}[h!]
\begin{center}
\includegraphics [height=6.2in,width=6.2in]{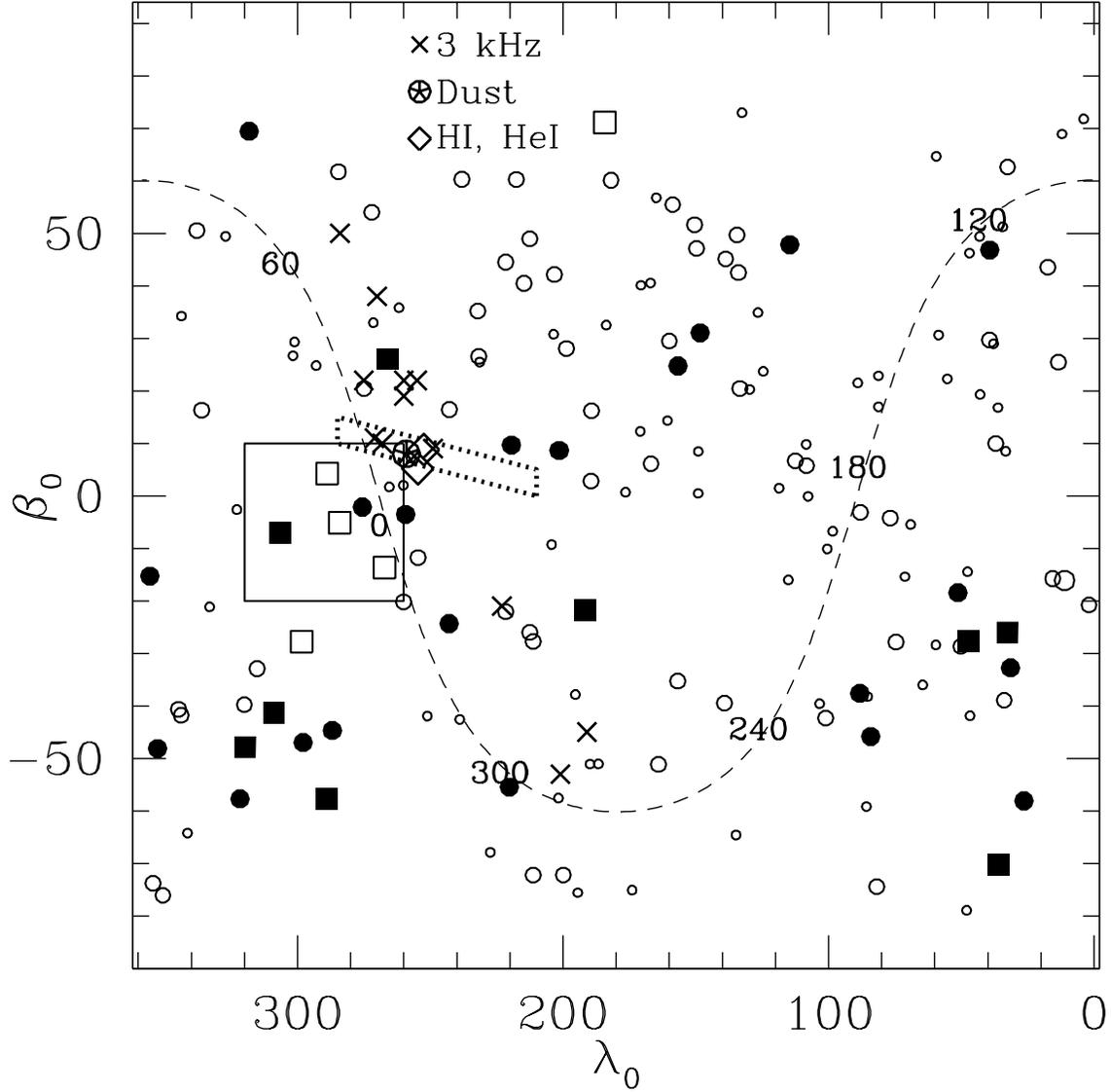}
\end{center}
\vspace{-0.1in}
\caption{The distribution of polarization is shown in ecliptic
coordinates.  
A region of polarization with similar position angles is
seen towards the heliosphere nose (solid line box, see Fig. \ref{fig:f2}
or Fig. 6 of T82).
A band of polarization extends at negative \eb 's from 
the heliosphere nose towards the downwind direction, which
is at \el$\sim$75\deeg, \eb$\sim$--5\deeg.  The upwind direction of inflowing dust grains, 
\HI, and \HeI\ are indicated by the circled
star and open diamonds.  The 3 kHz emission event locations are
plotted as ``X's''.  The dotted-line box shows the uncertainties on
the upwind dust direction \citep{Frischetal:1999}.  The solid-line box
shows the region of maximum \P\ from Figure 2.  The galactic plane is
shown as the dashed line labeled by galactic longitude.  The boxes and
circles represent stars with polarization \P$> 3 \sigma$ and \P$ = 2
\sigma- 3 \sigma$, respectively.  The filled symbols represent stars
within 30 pc, and the open symbols show stars with distances 30--40
pc.  The large and small open circles indicate, respectively, \P$ = 1
\sigma- 2 \sigma$ and \P$ < 1 \sigma$.  \label{fig:f3} }
\end{figure}

The offset of \Pmax\ corresponds to the secondary peak in energetic
neutral atom (ENA) fluxes found by \citet{Collieretal:2004}.  ENAs
originate in charge exchange between the plasma component and
interstellar \HI, which will happen inside of the orderly field
region that is required by the aligned grains.  However, since an
orderly \Bis\ will pinch off the plasma as it compresses against the
heliospheric magnetic field, it is understandable that the \Pmax\ and
ENA positions overlap (see Fig. \ref{fig:f2}).  Collier, however
(private communication), has suggested that ENA's may form by charge
exchange between \HI\ and the grains.  Since most of the surface area
of the ISDGs are in the small grain component that is excluded from
the heliosphere, this also is not unreasonable.

Finally, based on this analysis, there are two broad possibilities
that emerge.  It appears that sensitive observations of the 
polarization of nearby stars over long time scales will provide 
a ground-based monitor of the region where the heliosphere interacts 
with the ISM.  The second possibility is that interstellar dust
interacting with the heliosphere provides a low-$l$ contaminant
of the cosmic microwave background emission and/or polarization
data.  For this second case, the symmetries of the dust-heliosphere
interaction are strongly broken by the solar cycle, for both
large and small grains (but for different reasons).  These
broken symmetries should be evident from comparisons between solar
minimum and solar maximum data, and between data acquired over
different solar magnetic polarities.

\acknowledgements This research has been supported by the NASA
the grants NAG5-13107 and NNG05GD36G to the University of Chicago.



\begin{deluxetable}{cccc}
\tablecolumns{4}
\tablewidth{0pc}
\tablecaption{Composition of Interstellar Dust in the Cloud feeding Gas and Dust into the Heliosphere\label{tab:dust}}
\tablehead{
\colhead{Element} & \colhead{LIC\tablenotemark{a}} & \colhead{Solar\tablenotemark{b}} & \colhead{LIC\tablenotemark{c}} \\
\colhead{} & \colhead{Gas} & \colhead{Composition} & \colhead{Dust} \\
\colhead{} & \colhead{PPM} & \colhead{PPM} & \colhead{PPM} \\
}
\startdata
C  & 263--275 & 288 & 13--25  \\
N  &  50.1 -- 52.5 & 79 & 26.5 -- 28.9  \\
O & 380 & 575 & 195  \\
Mg & 4.68 -- 4.90 & 41.7 &  36.8 -- 37.0  \\
Al & 0.0794 & 3.47 & 3.39  \\
Si & 6.03 & 40.7 & 34.7  \\
Fe & 1.86 & 34.7 & 32.8  \\
\enddata

\tablenotetext{a}{The LIC gas-phase densities are from radiative transfer Models 2 and 8 in \citet{SlavinFrisch:2002}, also see \citet{FrischSlavin:2003}.}
\tablenotetext{b}{The solar composition is from \citet{Lodders:2003}.}
\tablenotetext{c}{The dust composition is given by the solar minus gas composition.}
\end{deluxetable}


\end{document}